\documentclass{emulateapj}

\shorttitle{V4743 Sgr as an intermediate polar }

\shortauthors{Kang et al.}

\begin{document}

\vskip 4 cm

\title {Nova V4743 Sagittarii 2002: An Intermediate Polar Candidate}
 
\author{Tae W. Kang$^{1}$, Alon Retter$^{1}$, Alex Liu$^{2}$, and Mercedes Richards$^{1}$}

\vskip 0.1 cm

\affil{$^{1}$Dept. of Astronomy \& Astrophysics, Penn State University, 525 Davey Lab, University Park, PA 16802; tkang@astro.psu.edu; retter@astro.psu.edu; mtr@astro.psu.edu}
\vskip 0.1 cm
\affil{$^{2}$Norcape Observatory, PO Box 300, Exmouth, 6707, Australia; asliu@onaustralia.com.au}
\vskip 0.1 cm

\begin{abstract}
We present the results of 11 nights of CCD unfiltered photometry of V4743 Sgr (Nova Sgr 2002 \# 3) from 2003 and 2005. We find two periods of 0.2799 d $\approx$ 6.7 h and 0.01642 d $\approx$ 24 min in the 2005 data. The long period is also present in the 2003 data, but only weak evidence of the shorter period is found in this year. The 24-min period is somewhat longer than the 22-min period, which was detected from X-ray observations. We suggest that the 6.7-h periodicity represents the orbital period of the underlying binary system and that the 24-min period is the beat periodicity between the orbital period and the X-ray period, which is presumably the spin period of the white dwarf. Thus, V4743 Sgr should be classified as an intermediate polar (DQ Her star). About six months after the nova outburst, the optical light curve of V4743 Sgr seemed to show quasi-periodic oscillations, which are typical of the transient phase in classical nova. Therefore, our results support the previous suggestion that the transition phase in novae may be related to intermediate polars.
\end{abstract}


\section{Introduction}

Novae are a subclass of cataclysmic variables (CVs). In these binary systems, matter from a companion star is accreted onto the surface of a white dwarf. When this material gets sufficiently hot and dense for the white dwarf to ignite a thermonuclear runaway, it blows off the hot burning layer at the surface. This event, which leads to a rapid increase in the luminosity of the binary system, is called a nova outburst. 

Nova V4743 Sgr ($\alpha_{2000.0}=19^h01^m09^s.38, \delta_{2000.0}= -22^o00'05''.9$) was discovered by Katsumi Haseda (2002) at V $\approx$ 5 on 2002 September 20. Kato et al. (2002) confirmed that the object is an Fe II-class nova and measured the FWHM of the H$\alpha$ emission line of 2400 km s$^{-1}$. From the long-term light curve of the American Association of Variable Star Observers (AAVSO), Nielbock \& Schmidtobreick (2003) calculated that the time required for a decline of three magnitudes from maximum is $t_3$ $\approx$ 15 d. Morgan, Ringwald, \& Prigge (2003) also measured $t_2$ $\approx$ 9 d and $t_3$ $\approx$ 16 d, which make V4743 Sgr a very fast nova according to Table 5.4 of Warner (1995).  
 
Wagner et al. (2003) reported $B$-band photometry of V4743 Sgr taken during three nights in 2003 July. They detected a sinusoidal modulation with a period of 0.281 $\pm$ 0.003 d (6.7 h) in the data, and interpreted it as the orbital period of the binary system. Ness et al. (2003) presented the results of two nights, 2003 April 4 and July 18, of Chandra X-ray observations of V4743 Sgr. They found a period of 1324 sec and suggested that it might represent either a pulsation or a spin period. Orio \& Tepedelenlioglu (2004) found several peaks in the power spectrum of V4743 Sgr from XMM-Newton observations, with the highest peak at a period of 1365 sec.  

Currently, there are about 50 novae with known orbital periods (Warner 2002). Finding the orbital period of a nova yields an estimate of the secondary mass (e.g., Smith \& Dhillon 1998). In addition, detecting several periodicities in novae can help in classifying the system into different groups of cataclysmic variables such as magnetic systems, intermediate polars, or  permanent superhump systems (e.g., Diaz \& Steiner 1989; Baptista et al. 1993; Retter, Leibowitz \& Ofek 1997; Patterson et al. 1997; Skillman et al. 1997; Patterson \& Warner 1998; Retter \& Leibowitz 1998; Retter, Leibowitz \& Kovo-Kariti 1998; Patterson 1999; Retter, Leibowitz \& Naylor 1999; Skillman et al 1999; Patterson 2001; Woudt \& Warner 2001; Lipkin et al. 2001; Patterson et al. 2002; Warner 2002; Woudt \& Warner 2002; Kang et al. 2006). This yields valuable information on the magnetic field of the white dwarf and the presence or absence of the accretion disk.
 
We have an ongoing program to observe novae with small telescopes to search for periodicities in their optical light curves. In this paper, we present photometric observations of V4743 Sgr, which suggests the presence of an orbital period and a beat period between the orbital and spin periods. We thus propose that V4743 Sgr is an intermediate polar system.

\section{Observations}

V4743 Sgr was observed during 5 nights on 2003 June and 6 nights on 2005 June. The observations included 82.2 hours and 2267 data points in total, and the nightly mean length was about 7.5 hours. Table 1 presents a summary of the schedule of the observations. The photometry was carried out with a 0.3-m f/6.3 telescope coupled to an SBIG ST7E CCD camera. The telescope is located in Exmouth, Western Australia. The full frame resolution is 765 x 510 pixels at 9 microns square. This camera is attached to an Optec f5 focal reducer giving an image field of view of 15 x 10 arcmin resulting in an image scale of 0.85 pixel per arcsec square. The range of seeing for the data was 2.5 -- 3 arcsec. The exposure times were between 30 and 60 sec every 120 sec, and no filter was used. Aperture photometry was used in the reduction, with an aperture size of 12 pixels. We estimated differential magnitudes with respect to the comparison star, ``C'' (Guide Star Catalog (GSC) 6294-2448) and the check star, ``K'' (GSC 6294-2481). Their recorded GSC magnitudes are V = 12.4 and 13.7, respectively. The magnitudes of the stars were derived from the SBIG CCDOPS software.

Figure 1 displays the long-term light curve of V4743~Sgr from outburst until 2005 June. The data were compiled from the AAVSO. By combining the data from the AAVSO with our data, we obtained 3291 individual points. The times of our observations are marked on the graph as well. Since our observations were taken with no filter, we compared our data with the AAVSO data around the same time, and subtracted 1 mag from our estimates to compensate for the difference between the visual and unfiltered data. The resulting light curve suggests that the fading of the nova was not smooth. Kato (2003) pointed out that quasi-periodic brightness oscillations with a peak-to-peak amplitude of up to 1 mag seemed to be present in the light curve between HJD 2452716 (2003 March) and 2452908 (2003 December). Kiyota, Kato \& Yamaoka (2004) also represented the long-term light curve of V4743 Sgr compiled by the Variable Stars Network (VSNET), which displays similar oscillations.

\begin{table}
\caption{The observations time table}
\begin{center}
\begin{tabular}{crccc}

UT       & Time of Start    & Run Time  & Points 	\\
(yymmdd) & (HJD--2452000)   &  (Hours)  & number        \\
                                                        \\
030604   & 795.1066         & 6.9 	& 192 		\\
030605   & 796.0732         & 7.8 	& 220 		\\
030609   & 800.0777         & 7.7 	& 204		\\
030610   & 801.0709         & 7.8 	& 219		\\
030611   & 802.0834         & 7.7 	& 209 		\\
050618   & 1540.0458        & 7.8 	& 223 		\\
050619   & 1541.0664        & 7.3 	& 203 		\\
050620   & 1542.0773        & 7.0 	& 194 		\\
050624   & 1546.0746        & 7.5 	& 202 		\\
050625   & 1547.0789        & 7.4 	& 197 		\\
050626   & 1548.0624        & 7.3 	& 204 		\\

\end{tabular}
\end{center}
\end{table}

\section{Data Analysis}

\subsection{The 2005 light curve}

We first discuss our 2005 data since two periodicities were clearly detected in this season. The bottom panel in Figure 2 shows the normalized power spectrum (Scargle 1982) of our raw data in 2005. It is dominated by two similar alias patterns around two frequencies, $f_1$ = 3.573 $d^{-1}$ and $f_2$ = 60.884 $d^{-1}$, which correspond to the periodicities $P_1$ = 0.2799 $\pm$ 0.0002 d (6.7 h) and $P_2$ = 0.016425 $\pm$ 0.00002 d (24 min). 

The power spectra of the raw data in 2005 zoomed into the frequency intervals around these two peaks are shown in Figures 3b and 4b, respectively. The power spectrum in Figure 3b clearly shows the $f_1$ frequency with its $\pm$ 1 $d^{-1}$ aliases. Figure 4b displays the $f_2$ peak with its~$\pm$~1~$d^{-1}$ aliases. 

\subsubsection{Tests}

To confirm that the $f_1$ and $f_2$ peaks are real, we performed several tests. First, we checked to see if the $f_1$ peak in the power spectrum was an artefact of the window function. This was done by assigning 1 to the times of observations of the 2005 data, and checking the resulting diagram. There was no evidence for significant power at the location of the peak. 

We also checked whether the power spectrum of the check star minus comparison star (K--C) data had a similar power and pattern near $f_1$. However, the K--C power spectrum did not show any strong power near this peak. The $f_1$ frequency can not be a result of color effects due to the nova being bluer than the comparison star. This is because $f_1$ is not an harmonic of 1 $d^{-1}$, and since there is no evidence for such a periodicity in the power spectrum of the hour angle data. To further check the significance of $f_1$, we used the bootstrap method. First, we scrambled the magnitude values, arbitrarily assigned them to the times of the observations, and calculated the corresponding power spectrum. We then found the power of the highest peak in the power spectrum in the frequency range 2--5 $d^{-1}$. Finally, we plotted the histogram of the highest peaks of the 1000 simulations. This suggested that the $f_1$ peak was about 59$\sigma$ significant. Thus, we confirmed that the peak was significant. The results were similar for the tests applied to the $f_2$ peak.

\begin{figure}
\epsscale{1} 
\plotone{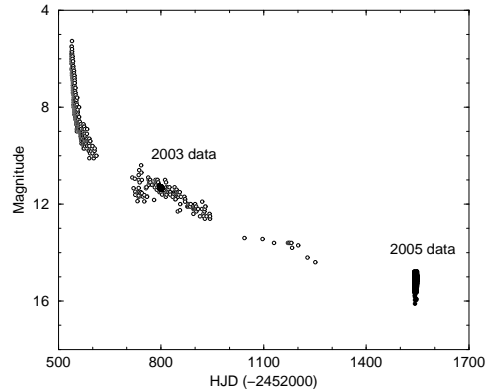}
\caption{The long-term light curve of V4743 Sgr. Empty circles represent visual estimates made by amateur astronomers, compiled by the AAVSO. The filled circles correspond to our observations. Our data points were shifted upwards by 1 mag to compensate for the difference between visual and unfiltered data. Quasi-periodic brightness oscillations with a peak-to-peak amplitude of up to 1 mag seemed to be present between HJD 2452716 (2003 March) and 2452908 (2003 December). Note that some variability could have been missed during the observational gap in 2002 November -- 2003 March.}
\vskip 0.3 cm
\end{figure}

\subsubsection{Significance of the two peaks}

To check the significance of two independent periodicities in the light curve of the nova, we applied a few tests. First, we fitted and subtracted the $f_2$ frequency from the data using a sinusoidal fit. The power spectrum of the residuals clearly showed $f_1$ as the dominant peak in the graph. Conversely, when the $f_1$ frequency was removed from the data, the $f_2$ frequency and its daily aliases remained in the residual power spectrum.

To check whether uncorrelated noise could be responsible for the presence of the candidate periods, we added noise to the model light curves of the 2005 data. The noise in the original data was defined as the root mean square of the data minus the $f_1$ frequency. We then searched for the highest peak in a frequency interval (55--65 $d^{-1}$) around the $f_2$ peak. In 1000 simulations, no peak reached the height of the $f_2$ frequency. Similarly, for the $f_1$ frequency, we did 1000 simulations in the frequency interval 2--5 $d^{-1}$. No peak reached the height of the $f_1$ peak either. So both periods seem to be real.

As a final test for the presence of the two frequencies, we divided our 2005 data into two parts (first and last three nights). In the power spectra of both parts of the observations, the same peaks appeared as the strongest ones in a 1 $d^{-1}$ frequency interval on both sides of the peaks (i.e., up to the 1 $d^{-1}$ aliases). Thus, we concluded that the two peaks in the 2005 data indicate real periodicities.  

In a search for a third frequency, we simultaneously fitted and subtracted the $f_1$ and $f_2$ frequencies from the data. The power spectrum of the residuals showed a few peaks in the 10--40 $d^{-1}$ frequency range. To analyze these peaks, we divided the 2005 data into two parts (first and last three nights). However, we did not see any consistent peak in the power spectra. Therefore, we believe that these peaks probably correspond to quasi-periodicities. 

\subsection{The 2003 light curve}

Figure 2a shows the normalized power spectrum of our 2003 data after subtracting the mean from the two parts of the data (first two and last three nights). This detrending method was used because the first two nights are about 0.07 mag brighter than the remaining nights. It displays the $f_1$ peak with its aliases as the dominant peak in the graph. 

The power spectra in 2003 zoomed into the frequency intervals 0--7 $d^{-1}$ and 57--65 $d^{-1}$ are presented in Figures 3a and 4a, respectively. Figure 3a clearly shows the $f_1$ frequency with its $\pm$ 1 $d^{-1}$ aliases. Figure 4a displays the $f_2$ peak with its $\pm$ 1 $d^{-1}$ aliases. However, we note that this frequency has a very low power.

We applied the same tests described in Section 3.1 to the 2003 data. The results of the $f_1$ peak were similar to the 2005 data. Thus, we confirmed that the $f_1$ frequency was also present in the 2003 data. However, the results of the $f_2$ peak were different. We concluded that this peak was not significant in our 2003 data.

In a search for a third frequency in the 2003 data, we fitted and subtracted the $f_1$ and $f_2$ frequencies from the data. The power spectrum of the residuals showed a few peaks in the 10--30 $d^{-1}$ frequency range. To check the significance of these peaks, we divided the 2003 data into two parts (first two and last three nights). However, we did not see any consistent peak in the power spectra.

\begin{figure}
\epsscale{1} 
\plotone{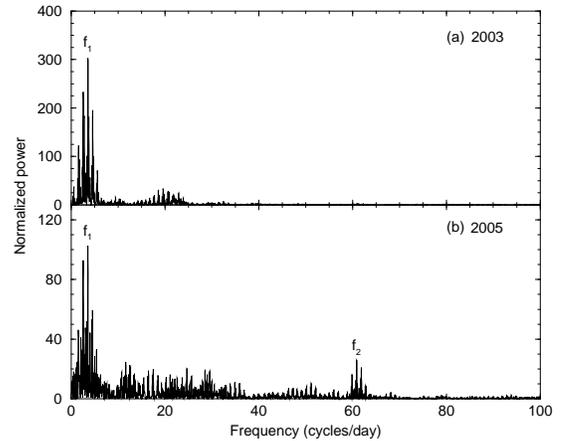}
\caption{Normalized power spectra of V4743 Sgr. (a) The power spectrum after removing the mean from the two parts of the data in 2003 (See Section 3.2). The $f_1$ = 3.573 $d^{-1}$ frequency is dominant, but the $f_2$  = 60.884 $d^{-1}$ peak cannot be seen because it is too weak. (b) The power spectrum of the raw light curve in 2005. Two frequencies, $f_1$ and $f_2$, are seen with their alias structures. See also Figures 3 and 4.}
\vskip 0.3 cm
\end{figure} 

\subsection{The structure of the periodicities}

In Figure 5 we present the light curve of V4743 Sgr folded on the 6.7-h period. The top panel shows the 2003 data after subtracting the mean from the two parts of the data and the bottom panel displays the raw data in 2005. Both are binned into 40 equal bins that cover the 0--1 phase interval. The peak-to-peak amplitudes of the mean variations were estimated as 0.065 $\pm$ 0.004 and 0.20 $\pm$ 0.01 mag for the 2003 and 2005 observations, respectively. The light curve of V4743 Sgr folded on the 24-min period is shown in Figure 6. The top panel displays the 2003 data after removing the mean from the two sets of the data and binned into 15 equal bins. The bottom panel presents the raw data in 2005 binned into 20 equal bins. The full amplitudes of the mean variations were 0.007 $\pm$ 0.003 and 0.10 $\pm$ 0.01 mag, respectively. The bars in Figures 5 and 6 are 1$\sigma$ uncertainties in the average values. The amplitudes of the mean variations were derived by fitting a sinusoidal curve to the mean light curve.

The best fitted ephemerides of the periodicities from the 2005 data were: 
\vskip 0.2 cm
$T_{1(min)}$(HJD) = 2453539.838(8) + 0.2799(2)E 
                
\vskip 0.2cm
$T_{2(min)}$(HJD) = 2453540.046(5) + 0.01642(2)E
\vskip 0.2 cm 

where E is an integer.

\vskip 0.2 cm

A noteworthy feature seen in Figure 5b is the dip in the mean light curve of the 6.7-h period at phase 1. Its reliability was checked by subtracting a pure sinusoidal with a peak-to-peak amplitude of 0.20 mag from the folded and binned data, resulting in a distribution of the points around zero level. This method gave an eclipse-like feature at phase 1 with a full amplitude of about 0.15 mag. To confirm the presence of the eclipses, we divided the 2005 data into the two parts (first and last three nights). We could clearly see an eclipse in the first three nights, but only found weak evidence of the eclipse during the last three nights. Therefore, this feature may be real.

\begin{figure}
\epsscale{1}
\plotone{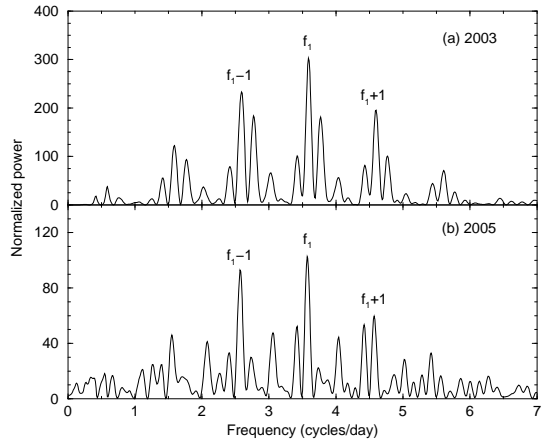}
\caption{The power spectra (Fig. 2) zoomed into the 0--7 $d^{-1}$ range of frequencies. They show the $f_1$ frequency with its $\pm$ 1 $d^{-1}$ aliases. (a) The power spectrum of the 2003 observations after subtracting the mean from the two sets of the data. (b) The power spectrum of the raw data in 2005.}
\vskip 0.3 cm
\end{figure}

\section{Discussion}
Our photometric data of V4743 Sgr show the presence of two independent periodicities in the light curve: $P_1$ = 6.7 h and $P_2$ = 24 min. We suggest that $P_1$ is the orbital period of the underlying binary system. Our conclusion is supported by the fact that this period was present in both light curves in 2003 and 2005, with no apparent change in its value. Wagner et al. (2003) also reported a detection of a 6.7 h periodicity in three nights of photometry in 2003 July using a B-filter, and interpreted this as the orbital period. 

The full amplitude of $P_1$ increased from 0.065 in 2003 to 0.2 mag in 2005 in our data. A similar behavior was observed in two other novae, V838 Her and V1494~Aql. Leibowitz et al. (1992) performed photometric observations of nova V838 Herculis 1991. They detected eclipses with a period of 7.1 h in the light curve and thus interpreted the periodicity as the obital period. They also found that the eclipse depth increased from about 0.1 mag in 1991 April to more than 0.4 mag in 1991 July. Bos et al. (2002) analyzed photometry of nova V1494 Aql 1999 \#2 during 12 nights in 2001. They found that the eclipse depth of the 3.2-h period was 10 times larger than in mid-2000. Both Leibowtiz et al. (1992) and Bos et al. (2002) interpreted the phenomenon as an eclipse of the accretion disk by the secondary star. Similarly, we propose to explain the 6.7-h variation in the light curve of V4743~Sgr as an eclipse of the accretion disk by the companion star or as a combination of an eclipse and aspect variations of the side of the comparison star, which is heated by the white dwarf (See also Balman, Retter \& Bos 2006). 

Within the errors, $P_2$ is the beat period between the orbital period and the spin period, which we take as the mean of the two values given for the X-ray period (Ness et al. 2003; Orio \& Tepedelenlioglu 2004): 1344 $\pm$ 21 sec $\approx$ 22 min. Ness et al. (2003) argued that the 22-min period could either be a pulsation or a spin period. However, if this periodicity represents the pulsation of the hot white dwarf, it should change as the temperature of the white dwarf cools down (e.g., Somer \& Naylor 1999). We estimated the ratio of the luminosity in 2003 April (when the X-ray period was detected) to 2005 June (when our observations ended). Using the relation between the luminosity and temperature of the white dwarf (Eq. 21 in Prialnik 1986), we estimated the change in the effective temperature. Then, we calculated the expected change in the frequency from the relation between frequency and temperature for stellar oscillations (Eq. 10 in Kjeldsen \& Bedding 1995). Thus, we estimated that, if interpreted as a pulsation, the 22-min period, which was found in 2003 April, should have increased to about 38 min in 2005 June. Since we did not detect any periodicity around 38 min, we believe that the 22-min period found in the X-ray band is the spin period, and not a pulsation period.

Wagner et al. (2003) reported continuous photometric observations of V4743 Sgr using a $B$-band during three nights in 2003 July. In addition to the presence of the long-term period of 6.7 h, they argued that another periodicity of 69 min existed in the data. Our observations could not confirm the presence of such a coherent period (Section 3), thus we believe that this is a quasi-periodicity.

\begin{figure}
\epsscale{1}
\plotone{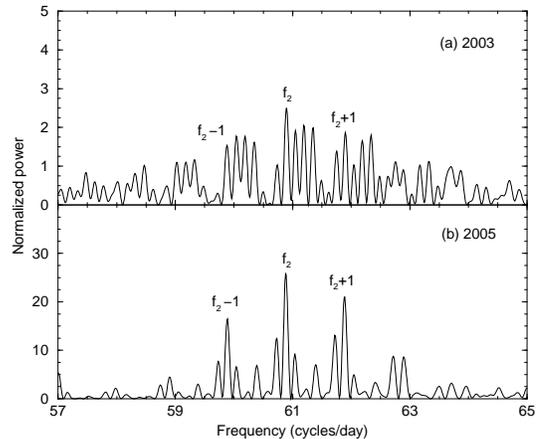}
\caption{The power spectra (Fig. 2) zoomed into the 57--65 $d^{-1}$ range of frequencies. They clearly show the $f_2$ frequency with its $\pm$ 1 $d^{-1}$ aliases in 2005, but in the 2003 data, this peak is very weak.}
\vskip 0.3 cm
\end{figure} 

\subsection{An Intermediate Polar model}

The AM Her stars (magnetic CVs) differ from the non-magnetic systems in two important qualitative respects: (1) a strong magnetic field on the primary star funnels the infalling gas onto one or two localized accretion shocks near the white dwarf's magnetic pole(s), where X-ray bremsstrahlung and polarized optical/infrared cyclotron emission arise; and (2) the white dwarf spin and binary orbital motions are locked in a rigid corotating geometry (e.g., Schmidt \& Stockman 1991).

Intermediate polars (also known as DQ her stars) are binary systems, which are a subclass of the AM Her stars. Unlike in other members of the AM Her class, in intermediate polars the rotation of the primary white dwarf is not synchronized with the orbital motion of the binary system. The spin periods found in intermediate polars are usually much shorter than their orbital periods (Patterson 1994; Hellier 1996).

One of the major observational characteristics of intermediate polars is the presence of multiple periodicities in the power spectra of their light curves, emanating from the non-synchronous rotation of the primary with the orbital revolution. In fact, this characteristic has become, together with the presence of X-ray modulations, a major criterion for membership in the intermediate polar group. 

We found several characteristics of the intermediate polars in V4743 Sgr. First, we detected several periodicities in the light curve. Second, the orbital period (6.7 h) is much longer than the spin period (22 min). Third, the spin period was found in X-ray observations. Fourth, the short term periodicity $f_2$ shows up about where it should be for an orbital sideband of the X-ray pulse (See above). Finally, the orbital period and the spin period roughly fit the relation $P_{spin}$ $\approx$ 0.1$P_{orb}$, which applies to many other known intermediate polars (Norton, Wynn, \& Somerscales 2004). We thus suggest that V4743 Sgr should be classified as an intermediate polar. 

\begin{figure}
\epsscale{1}
\plotone{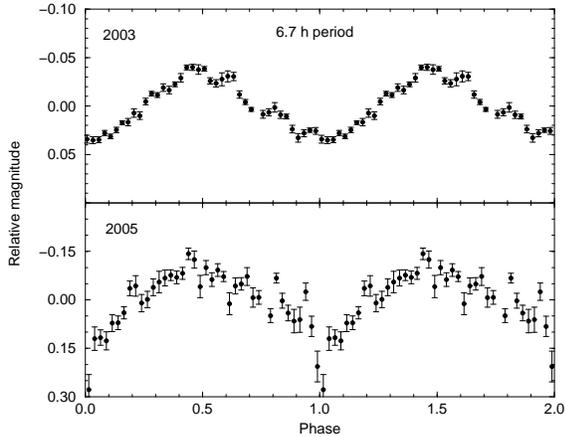}
\caption{The light curves of V4743 Sgr obtained in 2003 (top panel) and 2005 (bottom panel) folded on the 6.7-h period and binned into 40 equal bins. Two cycles are shown for clarity. Note the different scale in the y-axis and the dip at phase 1 in the bottom panel, which may indicate an eclipse.}
\vskip 0.3 cm
\end{figure}

\subsection{The Transition Phase in novae}

The optical light curves of classical novae are typically characterized by a smooth decline. However,  certain novae show a deep minimum in the light curve while others have slow oscillations with amplitudes of 1--2 mag typically a few weeks--months after maximum, during the so called `transition phase'. The deep minimum (e.g., in DQ Her 1934 -- Walker 1958) is understood by the formation of an optically thick dust envelope around the binary system. Several models for the oscillations during the transition phase have been proposed: oscillations of the common envelope, dwarf-nova outbursts, pulsations of the hot white dwarf, and winds, but there is still not enough observational evidence to understand this phenomenon (see Retter 2002 for a review). Retter (2002) argued that the explanation by oscillations of the common envelope can be easily rejected since this phase lasts only about 1--2 days after the nova event. He further concluded that dwarf nova outbursts can be ruled out as well because the accretion disks in young post-novae are thermally stable (e.g., Retter \& Naylor 2000; Schreiber, Gansicke, \& Cannizzo 2000). Shaviv (2001) proposed that steady-state winds of systems with super Eddington luminosities can be a natural explanation for the transition phase, and showed that a steady-state super Eddington stage could be maintained in a few novae.

Retter, Liller, \& Gerradd (2000b) observed LZ Mus, a nova that erupted in 1998, and clearly showed oscillations during the transition phase. They found several periods in its light curve. Therefore, they suggested that LZ~Mus is an intermediate polar system. They further proposed a new solution to the transition phase in novae by invoking a possible connection between the transition phase and intermediate polars. It was also found that LZ Mus was a strong X-ray source in 2001, which is consistent with the idea that the white dwarf has a strong magnetic field (Hernanz \& Sala 2002). 

Retter (2002) argued that the observations are consistent with the proposed link between the transition phase in novae and intermediate polars. For examples, nova GK Per 1901 showed transition phase oscillations. Its orbital period was found in optical observations and the spin period was detected in X-ray observations (Bianchini \& Sabadin 1983). V603 Aql 1918 was another transition-oscillation nova. It was suggested several times as an intermediate polar and showed strong variations in its X-ray light curve. However, there has been an extended discussion whether this nova is indeed an intermediate polar (e.g., Borczyk, Schwarzenberg-Czerny, \& Szkody 2003).

\begin{figure}
\epsscale{1}
\plotone{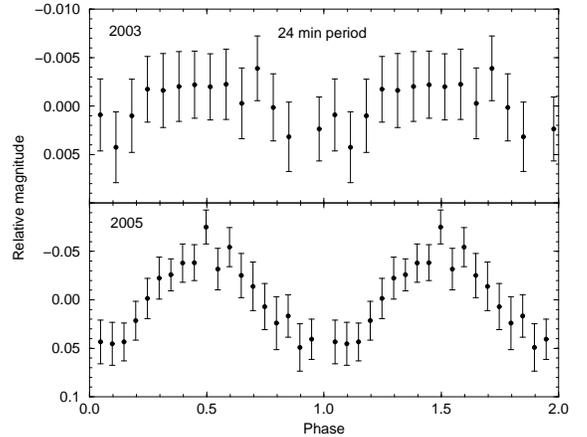}
\caption{The light curves of V4743 Sgr obtained in 2003 (top panel) and 2005 (bottom panel) folded on the 24-min period and binned into 15 equal bins for the 2003 data and 20 equal bins for the 2005 data. Two cycles are shown for clarity. Note the different scale in the y-axis.}
\vskip 0.3 cm
\end{figure} 

Retter (2002) pointed out that about 10$\%$ of the cataclysmic variable population are classified as intermediate polars, which is consistent with the rarity ($\sim15\%$) of the transition phase in novae. Note that these values are low limits due to the lack of observations. He hypothesized that the accretion disks in intermediate polars are easy to destroy in the nova outburst unlike other systems because the mass of the accretion disk in intermediate polars is much lower than in non-magnetic CVs. Observational evidence for the presence of the accretion disk was found in a few novae several weeks--months after eruption (e.g., Leibowitz et al. 1992; Skillman et al. 1997; Retter et al. 1997, 1998; Retter 1999, Iijima \& Esenoglu 2003; Retter 2004; Kang et al 2006). The fact that the accretion disk is recovered around the same time the transition phase occurs supports a connection between the two phenomena. 

Cs\'ak et al. (2005) found that spectra obtained during the maxima of the oscillations in the transition phase of V4745 Sgr are similar to spectra taken just after its nova eruption in 2003. Therefore, they proposed to explain the oscillations by several episodes of mass ejection, which they called mini nova-outbursts. This can be understood by the repeated destruction of the accretion disk and its re-establishment because of high mass transfer rates. The material accumulated on the surface of the white dwarf heats it and then another mini nova-outburst occurs. We note that the critical mass required for the mini-outburst is lower than the nova event because the white dwarf is hotter than before eruption.  

The proposed link between the transition phase in novae and intermediate polars (Retter et al. 2000b; Retter 2002) makes a simple prediction: novae whose light curves show transition phase oscillations should be intermediate polars. Iijima \& Esenoglu (2003) presented a spectroscopic coverage of the transition phase in nova V1494 Aql 1999 and displayed its optical light curve. Its transition stage started a month after maximum brightness (Kiss \& Thompson 2000), and the amplitude of the oscillations during the transition phase was $\sim$1 mag similar to V4743 Sgr (see Fig. 1). A period of 3.2 h was discovered in its optical light curve (Retter et al. 2000a) and was interpreted as the orbital period of the binary system. Drake et al. (2003) detected a 2523 sec periodicity in two X-ray runs using Chandra. Therefore, V1494 Aql may be an intermediate polar, but Drake et al. (2003) interpreted this period as the pulsation period of the white dwarf.

V1039 Cen 2001 is another nova that showed oscillations in its optical light curve during the transition phase (Kiyota, Kato \& Yamaoka 2004). Woudt, Warner \& Spark (2005) found two periodicities, 5.92 h and 719 s, in its light curve, which they respectively interpreted as the orbital and spin periods. Therefore, they concluded that the transition-oscillation nova V1039 Cen is an intermediate polar system as well.  

Two more examples of recent transition-oscillation novae are V2540 Oph 2002 (Ak, Retter, \& Liu 2005) and V4745~Sgr 2003 (Cs\'ak et al. 2005). In both cases, the orbital periods were found from optical photometry (Ak et al. 2005), but no short-term periodicities were detected. To our knowledge, no X-ray observations of these two novae were made. Another example of an old transition-oscillation nova is V373 Sct (Mattei 1975). In the light curve of V373 Sct, Woudt \& Warner (2003) detected a period of 258 sec, which is very likely the spin period of the white dwarf, but no long-term period was found. Thus, the intermediate polar model for V2540 Oph, V4745 Sgr and V373 Sct has yet to be confirmed. 

\section{Summary and Conclusion}

(1) We found two periods of 0.2799 d $\approx$ 6.7 h and 0.01642 d $\approx$ 24 min in the 2005 photometric observations of V4743 Sgr. In the 2003 data, we also detected the longer periodicity, but only found weak evidence of the shorter period.

(2) We interpret the 6.7-h period as the orbital period of the binary system and the 24-min period as the beat periodicity between the orbital period and the spin period, which was found in X-ray observations. We thus suggest that V4743 Sgr should be classified as an intermediate polar.

(3) The light curve of V4743 Sgr seemed to show quasi-periodic oscillations that are typical of the transition phase in novae. Thus, our results support the previously suggested link between transition-oscillation novae and intermediate polars.

(4) In three cases (V1494 Aql, V1039 Cen and V4743 Sgr) the proposed connection between transition-oscillation novae and intermediate polars seems to be vaild. However, more observations in the X-ray and optical bands are required to confirm or refute the proposed link, and to follow the evolution in time of the periodicities in V4743 Sgr.

\vskip 0.3 cm

We thank the anonymous referee for very useful comments that improved the paper. We acknowledge the amateur observers of the AAVSO who made the observations that comprise the long-term light curve of nova V4743 Sgr used in this study. TWK was supported by a REU supplement to NSF grant AST -- 0434234 (PI -- G. J. Babu).  AR was partially supported by a research associate fellowship from Penn State University.

\end{document}